\documentclass[twocolumn,preprintnumbers,prd,superscriptaddress,groupedaddress]{revtex4}
\usepackage{lmodern}
\usepackage[T1]{fontenc}
\usepackage{inputenc,psfrag,graphicx,epsfig,amsmath,amssymb,amsfonts,bbold,bm,manfnt,color,feynmp,caption,makeidx,hyperref}
\def\beq{\begin{equation}}
\def\eeq{\end{equation}}
\def\beqa{\begin{eqnarray}}
\def\eeqa{\end{eqnarray}}

\newcommand{\tildeJ}{\widetilde J}
\newcommand{\plainJ}{J}

\def\eps{\epsilon}

\newcommand{\refE}[1]{eq.~(\ref{#1})}

\begin{document}

\preprint{CERN-PH-TH-2017-056, CP3-17-06, Edinburgh 2017/06, FR-PHENO-2017-005} 

\title{\boldmath
The algebraic structure of cut Feynman integrals and the diagrammatic coaction
\unboldmath}

\author{Samuel Abreu}
\affiliation{Institute of Physics, University of Freiburg, D-79104 Freiburg, Germany}
\author{Ruth Britto}
\affiliation{School of Mathematics, Trinity College, Dublin 2, Ireland}
\affiliation{Hamilton Mathematics Institute, Trinity College, Dublin 2, Ireland}
\affiliation{Institut de Physique Th{\'e}orique, Universit\'e Paris Saclay, 
CEA, CNRS, F-91191 Gif-sur-Yvette cedex, France}

\author{Claude Duhr}
\affiliation{CERN Theory Division, 1211 Geneva 23, Switzerland}
\affiliation{Center for Cosmology, Particle Physics and Phenomenology (CP3),
Universit\'{e} Catholique de Louvain, 1348 Louvain-La-Neuve, Belgium} 

\author{Einan Gardi} 
\affiliation{Higgs Centre for Theoretical Physics, 
School of Physics and Astronomy, \\
The University of Edinburgh, Edinburgh EH9 3FD, Scotland, UK}

\begin{abstract}

\noindent
We study the algebraic and analytic structure of Feynman integrals by proposing an operation that maps an integral into pairs of integrals obtained from a master integrand and a corresponding master contour. This operation is a coaction. It reduces to the known coaction on multiple polylogarithms, but applies more generally, e.g.~to hypergeometric functions. The coaction also applies to generic one-loop Feynman integrals with any configuration of internal and external masses, and in dimensional regularization. In this case, we demonstrate that it can be given a diagrammatic representation purely in terms of operations on graphs, namely contractions and cuts of edges.
The coaction gives direct access to (iterated) discontinuities of Feynman integrals and facilitates a straightforward derivation of the differential equations they admit. In particular, the differential equations for any one-loop integral are determined by the diagrammatic coaction using limited information about their maximal, next-to-maximal, and next-to-next-to-maximal cuts.  

\end{abstract}


\maketitle

\noindent

Feynman integrals are central to perturbative quantum field theory (QFT), and it was realized early on that their analytic structure and discontinuities are directly connected to the fundamental concept of unitarity~\cite{Landau:1959fi,Cutkosky:1960sp}. Despite their importance in a broad range of applications over decades, the explicit computation of Feynman integrals is still difficult, sometimes prohibitively so. Controlling the analytic structure of Feynman integrals is key to precision collider physics, which requires {fast} evaluation of scattering amplitudes with an increasing number of loops and legs.  Developing {a} better understanding of the analytic structure of Feynman integrals can both help in devising new methods of computing them and shed light on fundamental aspects of QFT.

Feynman integrals are not only fundamental to QFT, but they are also deeply connected to certain areas of modern mathematics. This connection, in turn, has recently led to major advances in precision computations in particle physics. Instrumental to this progress was the realization that large classes of Feynman integrals can be evaluated in terms of so-called multiple polylogarithms (MPLs)~\cite{GoncharovMixedTate,Goncharov:2005sla}, a class of special functions that generalize the classical logarithm and polylogarithms to many variables. Understanding the mathematics of MPLs and their algebraic structure has led, for example, to new ways of evaluating certain classes of integrals in an algorithmic way~\cite{Brown:2008um,Anastasiou:2013srw,Panzer:2014caa,Bogner:2014mha,Bogner:2015nda,Ablinger:2014yaa}, to efficient approaches to solving differential equations~\cite{Henn:2013pwa}, and to {drastic simplifications of} complicated analytical results for Feynman integrals~\cite{Goncharov:2010jf}. 
An important aspect of MPLs is that they can be endowed with a so-called coaction~\cite{Remiddi:1999ew,GoncharovMixedTate,Brown:2009qja,Brown:2011ik,Duhr:2011zq,Duhr:2012fh}, an operation which maps a given MPL into pairs of simpler MPLs, effectively capturing the algebraic and analytic complexity of these functions.

However, not every Feynman integral can be evaluated in terms of MPLs. Indeed, generalizations to elliptic curves arise beyond one loop~\cite{Caffo:1998du,Bloch:2013tra,Adams:2013kgc,Bloch:2014qca,Adams:2014vja,Adams:2015gva,Adams:2015ydq,Bloch:2016izu,Adams:2016xah,Remiddi:2016gno,Primo:2016ebd,vonManteuffel:2017hms},
and the space of relevant functions has yet to be fully explored.
It is therefore important to investigate the algebraic structure of Feynman integrals, and the functions they evaluate to, beyond the context of MPLs.
In this letter, we take a step in this direction: we propose a coaction that generalizes the one on MPLs to a larger class of integrals and we explore its consequences for one-loop Feynman integrals in dimensional regularization.
In this case, our proposed coaction can be cast in a remarkably elegant form through simple operations on graphs. We give sample applications of our coaction to the study of Feynman integrals. In particular, we show how to obtain a compact representation of the complete set of differential equations satisfied by one-loop integrals with an arbitrary number of scales.\\

\noindent {\bf A coaction on integrals.} Consider a differential form $\omega$ which we assume to be closed (so that $\omega$ defines a cohomology class), and consider a contour $\gamma$ such that the integral of $\omega$ over $\gamma$ converges. We propose  the following coaction on such integrals:
\begin{equation}
\label{coaction}
\Delta\left(\int_\gamma \omega\right) = \sum_i \int_{\gamma} 
\omega_i  \otimes \int_{\gamma_i} \omega\,,
\end{equation}
where the sum runs over a basis $\omega_i$ of differential forms, called \emph{master integrands}. The contours $\gamma_i$ are called \emph{master contours}, and they are dual to the basis elements $\omega_i$ in the following sense:
\begin{equation}
\label{dual-def}
P\!_{ss}\!\left(
\int_{\gamma_i}\omega_j
\right)= \delta_{ij}\,,
\end{equation}
where $P\!_{ss}$ denotes a projector onto the subspace of `semi-simple' objects, defined as those objects $x$ on which the coaction acts trivially as $\Delta(x) = x\otimes 1$. Semi-simple objects include all rational and algebraic functions~\cite{Brown:motivicperiods}, certain transcendental numbers such as $2\pi i$~\cite{Brown:2011ik} and complete elliptic integrals~\cite{BrownTalk}. In contrast, examples of non-semi-simple objects include the classical logarithm and polylogarithm functions,  except where they evaluate to powers of $2\pi i$. The second integral on the right-hand side of eq.~\eqref{coaction} is defined only modulo transcendental semi-simple objects. We have checked that with these definitions, eq.~\eqref{coaction} satisfies the axioms of a coaction.  In particular, while individual terms on its right-hand side depend explicitly on a choice of master integrands, their sum is independent of this choice.

One can verify that eq.~\eqref{coaction} reduces to the coaction on MPLs \cite{GoncharovMixedTate,Brown:2011ik} when acting on this class of functions. In this case, the master integrands $\omega_i$ are differential forms with a subset of the poles of $\omega$. 
For each $\omega_i$, we define a corresponding master contour $\gamma_i$ by encircling the same subset of poles and dividing by a factor of $2\pi i$ for each pole. Hence the right entries in eq.~(\ref{coaction}) are obtained by taking residues of $\omega$ at the selected poles. 
The coaction of eq.~\eqref{coaction}  generalizes the one on MPLs to a larger class of functions while preserving some of its most important properties. In particular, it interacts with discontinuities and differential operators in the same way as the MPL coaction~\cite{Duhr:2012fh,Brown:motivicperiods}:
\begin{equation}\label{eq:discDiff}
\Delta\textrm{Disc} = (\textrm{Disc}\otimes\textrm{id})\Delta\quad
\textrm{and}\quad
\Delta\partial = (\textrm{id}\otimes \partial)\Delta\,,
\end{equation}
where $\textrm{id}$ denotes the identity operator.
As an example, eq.~\eqref{coaction} predicts the coaction on certain classes of hypergeometric functions. In the simple case of a Gauss hypergeometric ${}_2F_1$ function, the master integrands are two independent solutions to the hypergeometric differential equation, while the corresponding master contours are two straight lines. In Feynman integral calculations, hypergeometric functions appear when computing integrals in dimensional regularization, and we are typically interested in their Laurent expansion around small values of the dimensional regulator $\eps$. In certain cases, the coefficients of the Laurent series are MPLs, see e.g.~ref.~\cite{Moch:2001zr,Moch:2005uc,Huber:2005yg,Huber:2007dx}, and we have checked that the prediction of eq.~\eqref{coaction} for the coaction prior to expansion reproduces the same result as acting with the MPL coaction on the Laurent coefficients. Details will be given in a forthcoming publication.

In the remainder of this letter, we explore the consequences of eq.~\eqref{coaction} for one-loop Feynman integrals in dimensional regularization. We demonstrate that in this case eq.~\eqref{coaction} has a purely diagrammatic interpretation: the coaction of any one-loop Feynman integral can be expressed in terms of other Feynman integrals. The latter correspond to graphs that are obtained from the original one through two types of graphical operations, namely \emph{contractions} and \emph{cuts} of its edges.\\

\noindent {\bf A coaction on one-loop (cut) integrals and graphs.}
All one-loop integrals with numerators can be reduced to scalar integrals with unit numerator. Integrals with different powers of the propagators are related by integration-by-parts (IBP) identities \cite{Tkachov:1981wb,Chetyrkin:1981qh}, and integrals in different spacetime dimensions by dimension-shift identities \cite{Bern:1992em,Tarasov:1996br,Lee:2009dh}. It is therefore sufficient to discuss a basis of one-loop integrals of the form
\begin{equation}
\label{integral}
\tildeJ_n=\frac{e^{\gamma_{E}\epsilon}}{i\pi^{{D}/{2}}}\int d^{D}\!k\prod^{n-1}_{j=0}\frac{1}{\left(k-q_j\right)^2-m_j^2}\,,
\end{equation}
where $\gamma_E=-\Gamma'(1)$ is the Euler-Mascheroni constant, $n$ is the number of propagators, and the number of spacetime dimensions is accordingly chosen to be $D=2\lceil\frac{n}{2}\rceil-2\eps$.  
It is expected that at one loop all Feynman integrals can be expressed in terms of MPLs, order by order in the dimensional regulator $\epsilon$. The integrals $\tildeJ_n$ are a convenient basis, as they are functions of uniform weight $\lceil\frac{n}{2}\rceil$ (where $\eps$ is counted with weight $-1$).
We  divide $\tildeJ_n$ by its maximal cut in integer dimensions to define a normalized integral $\plainJ_n$.

A {\em cut integral} {$\mathcal{C}_C\plainJ_n$} in dimensional regularization~\cite{Abreu:2017ptx} is defined  from the uncut integrals of eq.~(\ref{integral}) by  integrating the integrand of {$\plainJ_n$} over a contour encircling the poles of a subset $C$ of propagators.
Each element $\plainJ_n$ of our basis is naturally represented by its  {\em Feynman graph}~$G$, with $n$ internal and $n$ external edges.  Each edge carries a distinct label, which can be used to keep track of its mass. We denote the set of all internal edges of $G$ by $E_G$. Similarly, we represent each cut integral $\mathcal{C}_CJ_G$ by a {\em cut graph} $(G,C)$, where $C$ is the subset of edges identifying the cut propagators, and where the original uncut integral $J_G\equiv J_n$ corresponds to $G\equiv(G,\emptyset)$.  

Within this setting, we find that when restricted to one-loop integrals, the coaction of eq.~\eqref{coaction} has a simple interpretation as a coaction on cut graphs. 
We first note that there is a combinatorial coaction on cut graphs \footnote{This is a version of an {\em incidence algebra \cite{JoniRota}}.} given by
\begin{equation} 
\label{diag-conj}
\Delta(G,C)=\!\!\!\!\!\!\!\!
\sum_{\substack{C\subseteq X\subseteq E_G\\X\neq \emptyset}}
\!\!\!\!\! \Big((G_X,C) +a_X\!\!\! \sum_{e\in X\setminus C}\!\! (G_{X\setminus e},C) \Big)\otimes (G,X),
\end{equation}
where $a_X$ can be any function on subsets of edges that takes values in $\mathbb{Q}$, and $G_X$ is the graph obtained by contracting all internal edges of $G$ except those in $X$. 
The second entry in the coaction is the cut graph $(G,X)$ where 
the set of cut edges $X$ is necessarily nonempty and contains the subset $C$.
A central result of our paper is that there exist unique values of $a_X$ 
for which the graphical operation of eq.~(\ref{diag-conj}) maps precisely to 
the coaction of eq.~\eqref{coaction} when acting on one-loop Feynman integrals.
To illustrate this, consider the application of eq.~\eqref{diag-conj} to the uncut basis integrals $J_G$ (i.e. taking $C=\emptyset$), yielding
 \begin{align}\label{eq:Delta_one-loop}
 	\Delta(J_G)=\!\!\sum_{\substack{\emptyset\neq X\subseteq E_G}}&
 	\left(J_{G_X}+a_X\sum_{e\in X}J_{G_X\setminus e}\right)\otimes \mathcal{C}_XJ_{G}\,.
 \end{align}
Equation~\eqref{eq:Delta_one-loop} agrees with the coaction of eq.~\eqref{coaction} on one-loop integrals, provided we set $a_X=1/2$ ($a_X=0$) when the number of edges $|X|$ is even (odd). 
This defines the operation we call the  diagrammatic coaction. Using 
eq.~(\ref{diag-conj}) it generalizes straightforwardly to any $\Delta({\cal C}_C J_G)$, with the same values for $a_X$.
To relate eq.~\eqref{coaction} to eq.~(\ref{eq:Delta_one-loop}), we establish that a complete set of contours is given by those encircling the poles of propagators, as in the cut integrals above, along with those that additionally encircle the pole at infinity~\cite{Abreu:2017ptx}.  These two types of contours are associated with Landau singularities of the first and second types, respectively. A remarkable linear relation~\cite{PhamCompact,Froissart,Teplitz}  then allows us to express the latter in terms of the former, so that it is indeed sufficient to write cut integrals $\mathcal{C}_XJ_{G}$ in the second entries of eq.~\eqref{coaction}. The same linear relation can be interpreted as generating the terms proportional to $a_X$, with its values fixed uniquely as above, in order to satisfy eq.~\eqref{dual-def}.

Considering a variety of one-loop integrals, we have verified that  eq.~\eqref{eq:Delta_one-loop} holds order by order in $\eps$. To this end, we computed the left-hand side by acting with the MPL coaction on the first few Laurent coefficients of a given integral $J_n$, while on the right-hand side we expanded  both the basis integrals and their cuts in $\epsilon$. In particular, we have verified  eq.~\eqref{eq:Delta_one-loop} for tadpoles, bubbles and most triangles and boxes with a variety of internal and external mass configurations~\cite{LongDiag}. We have also performed consistency checks for the massless pentagon and hexagon.  This leads us to conjecture that eq.~\eqref{eq:Delta_one-loop} holds to any order in $\epsilon$ for all basis integrals $J_n$, and hence for any one-loop integral. 

We present some illustrative examples with few edges. For the uncut bubble with massive propagators, we have
\begin{align}\label{eq:bub}
&\Delta\Big[\!\!\raisebox{-3mm}{
\includegraphics[keepaspectratio=true, width=1.3cm]{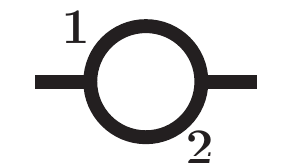}}\Big]
=\raisebox{-3mm}{\includegraphics[keepaspectratio=true, height=.8cm]{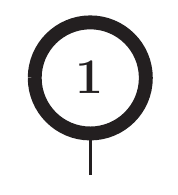}}\otimes
\raisebox{-3mm}{\includegraphics[keepaspectratio=true, width=1.3cm]{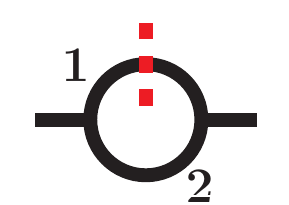}}
+\raisebox{-3mm}{\includegraphics[keepaspectratio=true, height=.8cm]{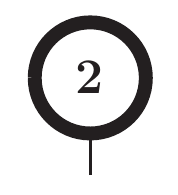}}\otimes
\raisebox{-3.5mm}{\includegraphics[keepaspectratio=true, width=1.3cm]{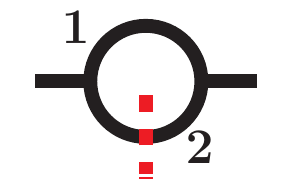}}
\nonumber 
\\
&+\left(\raisebox{-3mm}{
\includegraphics[keepaspectratio=true, width=1.3cm]{./diagrams/bubPRL.pdf}}+\frac{1}{2}\raisebox{-3mm}{\includegraphics[keepaspectratio=true, height=.8cm]{./diagrams/tadPRL1.pdf}}+\frac{1}{2}\raisebox{-3mm}{\includegraphics[keepaspectratio=true, height=.8cm]{./diagrams/tadPRL2.pdf}}\right)\otimes\raisebox{-3.5mm}{\includegraphics[keepaspectratio=true, width=1.3cm]{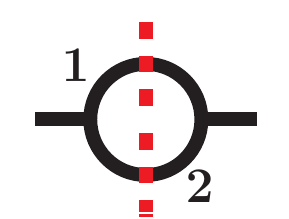}}\,.
\end{align}
For its two-propagator cut, we obtain
\begin{align}
\label{eq:cutbub}
\Delta \bigg[\raisebox{-3.5mm}{\includegraphics[keepaspectratio=true, width=1.3cm]{./diagrams/bubPRLCut12.pdf}}\bigg]
&=\raisebox{-3.5mm}{\includegraphics[keepaspectratio=true, width=1.3cm]{./diagrams/bubPRLCut12.pdf}}\otimes\raisebox{-3.5mm}{\includegraphics[keepaspectratio=true, width=1.3cm]{./diagrams/bubPRLCut12.pdf}}\,.
\end{align}
These equations can be read in two ways: according to~\refE{diag-conj}, as an operation on graphs, or according to eq.~(\ref{eq:Delta_one-loop}), as an operation on the functions obtained by evaluating the corresponding integrals $\plainJ_n$, as defined in eq.~(\ref{integral}).
Equation (\ref{eq:cutbub}) illustrates the generalization of eq.~(\ref{eq:Delta_one-loop}) to $\Delta\left({\cal C}_C J_n\right)$.
Since the formula for the coaction is valid before expansion in $\eps$, masses may be assigned arbitrarily. In some limits some of the cut integrals can vanish or contain divergences. These limiting behaviors do not spoil the validity of our conjecture. For instance, the coaction on a triangle with massless propagators (indicated by thin edges) is given by 
\begin{align}\label{eq:triangle}
&\Delta\left[\raisebox{-3.7mm}{\includegraphics[keepaspectratio=true, width=1.25cm]{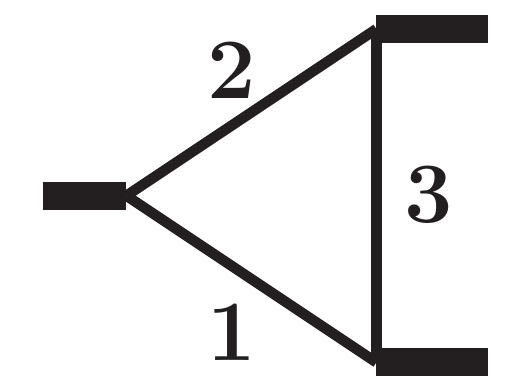}}\right]=
\raisebox{-2.7mm}{\includegraphics[keepaspectratio=true, width=1.2cm]{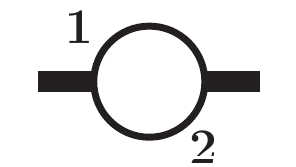}}\otimes\raisebox{-3.7mm}{\includegraphics[keepaspectratio=true, width=1.25cm]{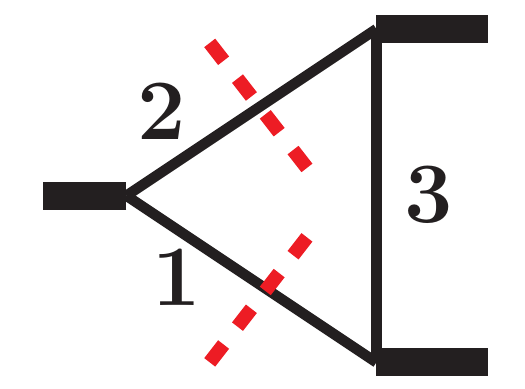}}
+\raisebox{-2.7mm}{\includegraphics[keepaspectratio=true, width=1.2cm]{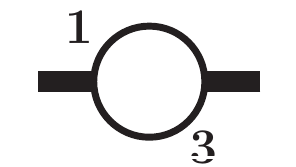}}\otimes\raisebox{-3.7mm}{\includegraphics[keepaspectratio=true, width=1.25cm]{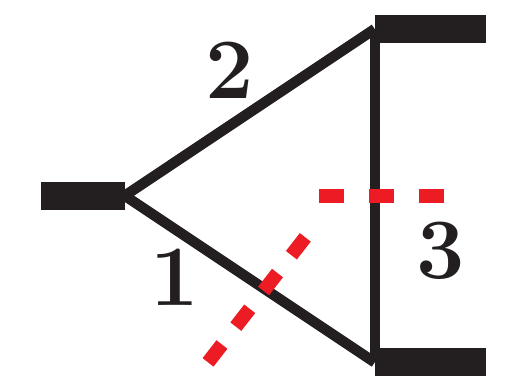}}\nonumber\\
&+\raisebox{-2.7mm}{\includegraphics[keepaspectratio=true, width=1.2cm]{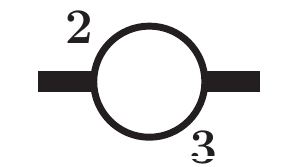}}\otimes\raisebox{-3.7mm}{\includegraphics[keepaspectratio=true, width=1.25cm]{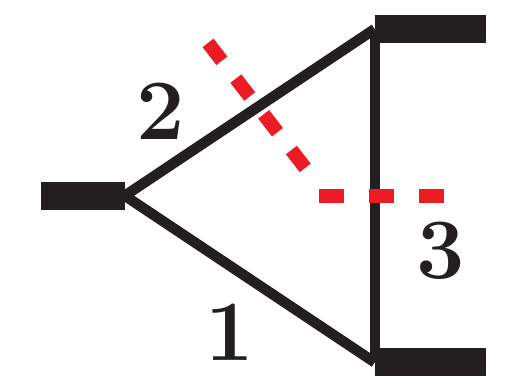}}
+\raisebox{-3.7mm}{\includegraphics[keepaspectratio=true, width=1.25cm]{./diagrams/triPRL.pdf}}\otimes\raisebox{-3.7mm}{\includegraphics[keepaspectratio=true, width=1.25cm]{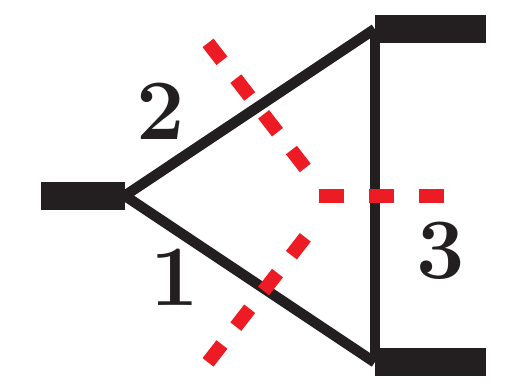}}\,\,,
\end{align}
since all tadpoles vanish in this case.
Given our choice of basis integrals in eq.~\eqref{integral} where $D=2\lceil\frac{n}{2}\rceil-2\eps$, the integrals corresponding to the uncut graphs on the left-hand sides of eqs.~\eqref{eq:bub} and~\eqref{eq:triangle} are finite, while the tadpoles in eq.~\eqref{eq:bub} and the massless bubbles in eq.~\eqref{eq:triangle} introduce singularities on the right-hand sides. Owing to a relation between uncut integrals and their one- and two-propagator cuts~\cite{Abreu:2017ptx}, these poles cancel, which is essential for the validity of eq.~(\ref{eq:Delta_one-loop}). This cancellation is a general feature of finite integrals and a nontrivial consistency check of our conjecture.\\

\noindent {\bf Discontinuities and the first-entry condition.}
The existence of a simple formula for the coaction on Feynman integrals is not only of formal interest.
One class of applications stems from the fact that Feynman integrals are multivalued functions whose discontinuities are described by cut integrals.
According to eq.~\eqref{eq:discDiff}, discontinuity operators  act only on the first entry of the coaction.
It has already been understood that the first entries in the coaction encode either branch cuts of external channels~\cite{Gaiotto:2011dt} or internal masses~\cite{Abreu:2015zaa}. This is known as the \emph{first-entry condition}. Equation~\eqref{eq:Delta_one-loop} implies a stronger version of this condition: the first entries of the coaction are themselves Feynman integrals. This generalizes a result of ref.~\cite{Brown:2015fyf} to divergent integrals in dimensional regularization and also incorporates the Steinmann relations~\cite{steinmann1,steinmann2,Cahill:1973qp,Caron-Huot:2016owq}.
The known version of the first-entry condition immediately follows.
The terms in the coaction with tadpoles and bubbles in the first entry identify branch cuts of internal masses and external channels, and the corresponding discontinuity functions appear in the respective second entries.
We refer to~\refE{eq:triangle} as an example: each bubble in the first entries of the tensor uniquely identifies a channel, and the corresponding discontinuity function, the relevant two-propagator cut, appears in the respective second entry. The relation between single discontinuities and cuts is thus satisfied by construction. Relations between iterated discontinuities and multiple cuts~\cite{Abreu:2015zaa,Abreu:2014cla} are also built in: for instance, double discontinuities are apparent in terms with triangles and boxes
in the first entry, and the corresponding three- and four-propagator cuts in the second.\\

\noindent {\bf Differential equations for one-loop integrals.} 
Another important application of eq.~(\ref{eq:Delta_one-loop}) is the possibility to easily derive differential equations for one-loop Feynman integrals. 
For this, we focus on the terms of eq.~(\ref{eq:Delta_one-loop}) with a logarithm in the second entry.
According to eq.~\eqref{eq:discDiff}, differential operators act only on the second entry of the coaction,
turning the logarithms into algebraic functions. 
The resulting tensor can thus be trivially lifted to a function.
As an example, consider the pentagon integral $J_5$ in $6-2\epsilon$ dimensions.
The terms on the right-hand side of eq.~(\ref{eq:Delta_one-loop}) with a
logarithm in the second entry are very few: they are the maximal cut of the pentagon at ${\cal O}(\epsilon^1)$, and the next-to-maximal and next-to-next-to-maximal cuts at ${\cal O}(\epsilon^0)$. From eq.~\eqref{eq:discDiff}, it follows that the derivative of $J_5$ is related exclusively to the Feynman integrals appearing in the first entries corresponding to these cuts, namely $J_5$, $J_4$ and $J_3$. Consequently, we obtain
\begin{widetext}
\begin{equation}
	\label{eq:pent}
	d\!\!\left[\raisebox{-5mm}{\includegraphics[keepaspectratio=true, width=1.3cm]{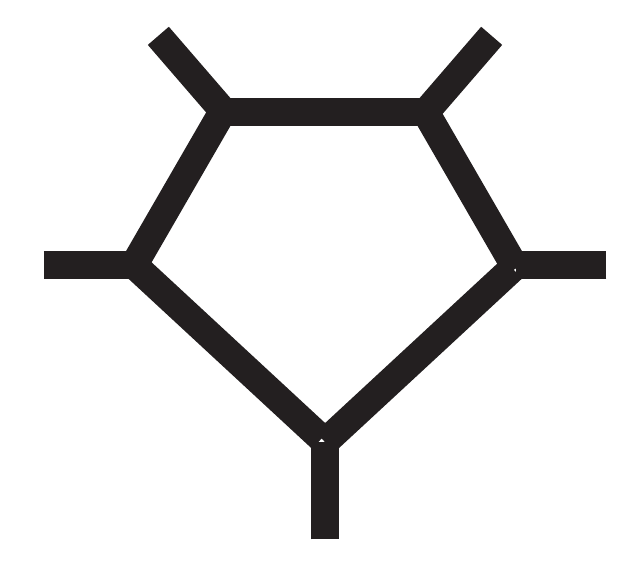}}\right]=
	\sum_{(ijk)}
\raisebox{-4mm}{\includegraphics[keepaspectratio=true, width=1cm]{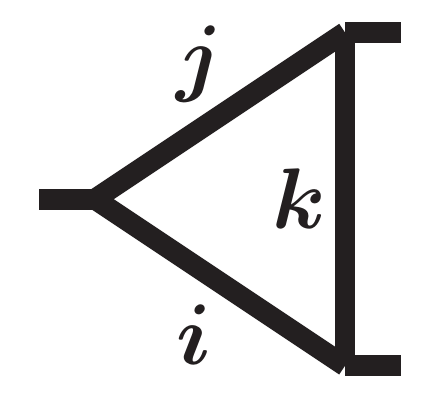}}d\!\!
\left[\raisebox{-5mm}{\includegraphics[keepaspectratio=true, width=1.3cm]{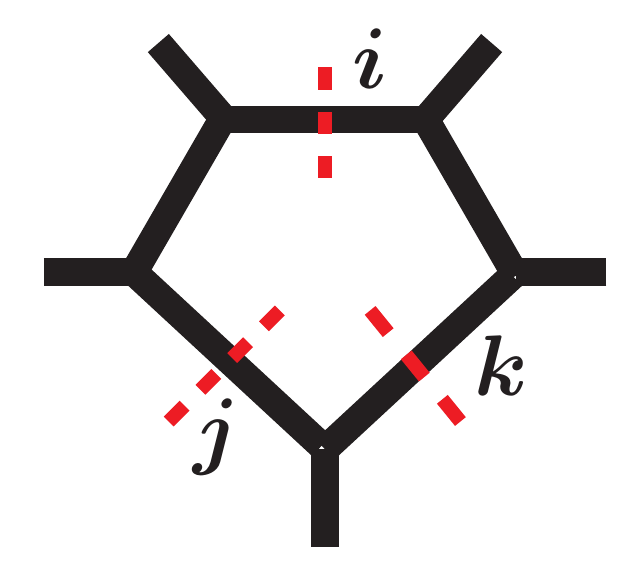}}\!+\!\frac{1}{2}\sum_{l}\raisebox{-5mm}{\includegraphics[keepaspectratio=true, width=1.3cm]{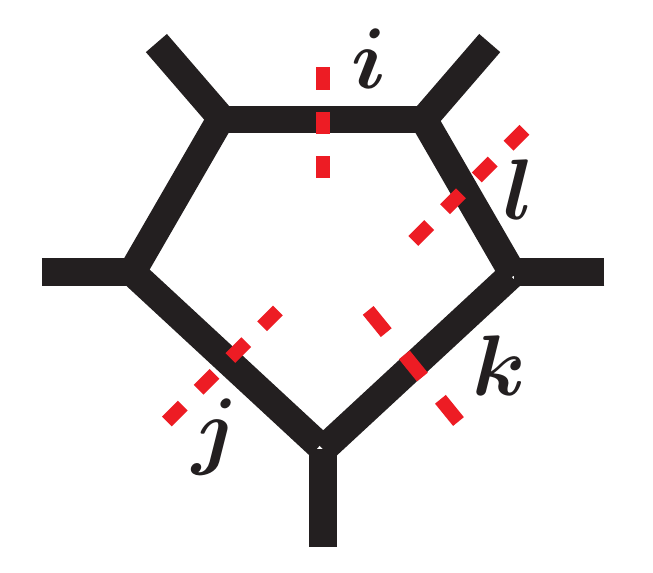}}\right]_{\epsilon^0}\!\!
+\!\sum_{(ijkl)}
	\raisebox{-4mm}{\includegraphics[keepaspectratio=true, width=1.3cm]{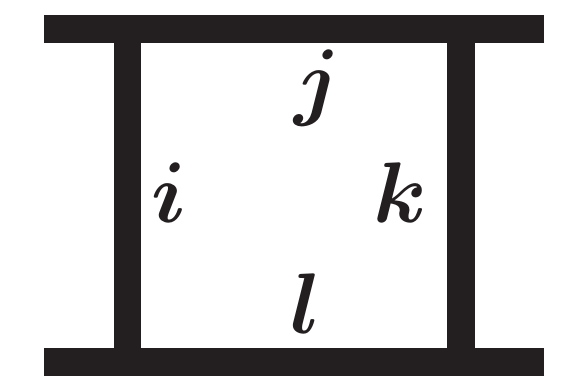}}
	d\!\!\left[\raisebox{-5mm}{\includegraphics[keepaspectratio=true, width=1.3cm]{./diagrams/pentagonTQuad2.pdf}}\right]_{\epsilon^0}
\!\!\!+\eps\,\raisebox{-5mm}{\includegraphics[keepaspectratio=true, width=1.3cm]{./diagrams/pentagon2.pdf}}
d\!\!\left[	\raisebox{-5mm}{\includegraphics[keepaspectratio=true, width=1.3cm]{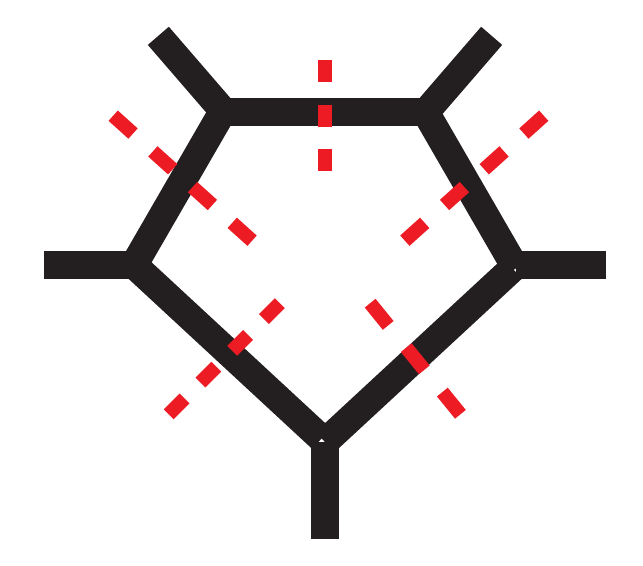}}\right]_{\epsilon^1},
\end{equation}
\end{widetext}
where $i,j,k$ and $l$ run over distinct edges of the graph.  We stress that the differential equation in eq.~(\ref{eq:pent}) is valid for a general pentagon integral, with any configuration of internal and external masses, and to all orders in $\epsilon$. 

Similarly, one may derive differential equations for any one-loop integral, with an arbitrary number of propagators $n$ and an arbitrary number of scales.  
It is a general feature that the arguments of the $d\log$ forms in these differential equations, also known as the \emph{letters} of the \emph{alphabet} of the integral, are associated exclusively with Feynman integrals $J_n$ with at most two uncut propagators. These cuts can be computed explicitly for any $n$ using the techniques of ref.~\cite{Abreu:2017ptx}. Hence, we are able to write down the system of differential equations for any one-loop integral.
We note that this system is characterized by a simple triangular structure: the equation for $J_n$ involves only $J_n$, $J_{n-1}$ and~$J_{n-2}$.
Details will be given in a forthcoming publication~\cite{LongDiag}.\\

\noindent {\bf Discussion.} In this letter we have proposed a coaction on certain classes of integrals. When acting on MPLs, it reduces to the known coaction on this class of functions. Our proposal, however, goes beyond this case, and we have demonstrated its validity by constructing a coaction on one-loop integrals that admit a Laurent expansion in terms of MPLs, reproducing the coaction on MPLs order by order in $\epsilon$. Remarkably, when restricted to one-loop integrals, eq.~\eqref{coaction} has a simple combinatorial interpretation in terms of cut graphs, given by eq.~\eqref{diag-conj}. 

The resulting diagrammatic coaction of eq.~(\ref{eq:Delta_one-loop}) effectively encodes the algebraic and analytic structure of one-loop integrals. It therefore has immediate applications in identifying the discontinuities of these integrals in terms of cuts, and in deriving the differential equations they admit.  These equations are determined by limited information on cut integrals. Standard methods to derive such differential equations rely instead on solving systems of IBP identities, which often becomes a computational bottleneck. Obtaining the differential equation from the diagrammatic coaction circumvents this difficulty. It also sheds light on the general structure of these systems of differential equations and the resulting alphabet. 

It is natural to speculate how our discussion might generalize beyond one loop, in particular to cases where the integrals can no longer be expressed in terms of MPLs. The concept of master contour has a natural interpretation in the context of two-loop integrals~\cite{CaronHuot:2012ab}, so it is reasonable to expect that eq.~\eqref{coaction} continues to hold if the appropriate master contours are considered. Starting from two loops, several master integrals may share the same propagators, and differ only by the powers to which propagators are raised or by polynomials in the numerator. In such cases we cannot expect that the master contours can only be labeled by subsets of propagators; rather, the contours will depend crucially on Landau singularities of the second type. As a consequence, while we believe that eq.~\eqref{coaction} continues to hold beyond one loop, it is still unclear how eq.~(\ref{eq:Delta_one-loop}) generalizes.
Integrations over more complicated classes of contours may be required, including integrations over cycles of surfaces of higher genus in cases where the integral cannot be expressed in terms of MPLs. The study of homology groups associated to certain two-loop integrals~\cite{Federbush} may help in identifying the relevant contours.

Finally, let us comment on connections to pure mathematics. First, the search for a combinatorial coaction on Feynman graphs which agrees with the coaction on MPLs has been an active area of research, see e.g.~ref.~\cite{Bloch:2005bh}. So far, however, coproducts and coactions on Feynman graphs have focused on Feynman graphs without cut edges~\cite{Kreimer:1997dp,Connes:1998qv,Connes:1999yr,Kreimer:2009jt}. 
Equation~(\ref{eq:Delta_one-loop}) suggests that it is in fact also necessary to take cut Feynman graphs into account. It would therefore be interesting to examine the  literature in the light of eq.~(\ref{eq:Delta_one-loop}).
Second, it would be interesting to compare eq.~\eqref{coaction} with the motivic coaction introduced by Brown~\cite{Brown:motivicperiods} and its application to Feynman integrals~\cite{Brown:2015fyf,Panzer:2016snt}. The motivic coaction takes a slightly different form. In particular, its second entry does not have a direct interpretation in terms of integrals, so its relation to eq.~\eqref{coaction} is not straightforward. Our results may shed new light on some of these concepts in pure mathematics, while the motivic coaction might elucidate the origin of the simple form of the diagrammatic coaction at one loop, paving the way for its generalization. We expect that eq.~\eqref{coaction} should apply equally well to two-loop integrals and hypergeometric functions that cannot be expressed in terms of MPLs, but require the introduction of elliptic curves. Equation~\eqref{coaction} thus makes a prediction for a coaction on elliptic generalizations of MPLs, and it would be interesting to explore this direction, and its connection to the motivic coaction.

\begin{acknowledgments}

We are grateful to Herbert Gangl, Roman Lee, Erik Panzer and Volodya Smirnov for discussions and communications, and to Francis Brown, Fabrizio Caola and Bernhard Mistlberger for comments on the manuscript.
SA acknowledges the hospitality of Trinity College Dublin and the CERN Theoretical Physics Department at various stages of this work.
CD acknowledges the hospitality of the Higgs Center of the University of Edinburgh and of Trinity College Dublin at various stages of this work. EG acknowledges the hospitality of Trinity College Dublin.  
This work is supported by the Juniorprofessor Program of Ministry of Science, Research and the Arts of the state of Baden-W\"urttemberg, Germany (SA), the ERC Consolidator Grant 647356 ``CutLoops'' (RB), the ERC Starting Grant 637019 ``MathAm'' (CD), and by the STFC Consolidated Grant ``Particle Physics at the Higgs Centre'' (EG). 
We would also like to thank the ESI institute in Vienna and the organizers of the program ``Challenges and Concepts for Field Theory and Applications in the Era of LHC Run-2'', Nordita in Stockholm and the organizers of ``Aspects of Amplitudes,'' over the summer of 2016, and the MITP in Mainz and the organizers of the program ``AMPDEV2017'', where certain ideas presented in this paper were consolidated.

\end{acknowledgments}

\bibliography{bibMain}

\begin{thebibliography}{62}%
\makeatletter
\providecommand \@ifxundefined [1]{%
 \@ifx{#1\undefined}
}%
\providecommand \@ifnum [1]{%
 \ifnum #1\expandafter \@firstoftwo
 \else \expandafter \@secondoftwo
 \fi
}%
\providecommand \@ifx [1]{%
 \ifx #1\expandafter \@firstoftwo
 \else \expandafter \@secondoftwo
 \fi
}%
\providecommand \natexlab [1]{#1}%
\providecommand \enquote  [1]{``#1''}%
\providecommand \bibnamefont  [1]{#1}%
\providecommand \bibfnamefont [1]{#1}%
\providecommand \citenamefont [1]{#1}%
\providecommand \href@noop [0]{\@secondoftwo}%
\providecommand \href [0]{\begingroup \@sanitize@url \@href}%
\providecommand \@href[1]{\@@startlink{#1}\@@href}%
\providecommand \@@href[1]{\endgroup#1\@@endlink}%
\providecommand \@sanitize@url [0]{\catcode `\\12\catcode `\$12\catcode
  `\&12\catcode `\#12\catcode `\^12\catcode `\_12\catcode `\%12\relax}%
\providecommand \@@startlink[1]{}%
\providecommand \@@endlink[0]{}%
\providecommand \url  [0]{\begingroup\@sanitize@url \@url }%
\providecommand \@url [1]{\endgroup\@href {#1}{\urlprefix }}%
\providecommand \urlprefix  [0]{URL }%
\providecommand \Eprint [0]{\href }%
\providecommand \doibase [0]{http://dx.doi.org/}%
\providecommand \selectlanguage [0]{\@gobble}%
\providecommand \bibinfo  [0]{\@secondoftwo}%
\providecommand \bibfield  [0]{\@secondoftwo}%
\providecommand \translation [1]{[#1]}%
\providecommand \BibitemOpen [0]{}%
\providecommand \bibitemStop [0]{}%
\providecommand \bibitemNoStop [0]{.\EOS\space}%
\providecommand \EOS [0]{\spacefactor3000\relax}%
\providecommand \BibitemShut  [1]{\csname bibitem#1\endcsname}%
\let\auto@bib@innerbib\@empty
\bibitem [{\citenamefont {Landau}(1959)}]{Landau:1959fi}%
  \BibitemOpen
  \bibfield  {author} {\bibinfo {author} {\bibfnamefont {L.}~\bibnamefont
  {Landau}},\ }\href {\doibase 10.1016/0029-5582(59)90154-3} {\bibfield
  {journal} {\bibinfo  {journal} {Nucl.Phys.}\ }\textbf {\bibinfo {volume}
  {13}},\ \bibinfo {pages} {181} (\bibinfo {year} {1959})}\BibitemShut
  {NoStop}%
\bibitem [{\citenamefont {Cutkosky}(1960)}]{Cutkosky:1960sp}%
  \BibitemOpen
  \bibfield  {author} {\bibinfo {author} {\bibfnamefont {R.}~\bibnamefont
  {Cutkosky}},\ }\href@noop {} {\bibfield  {journal} {\bibinfo  {journal}
  {J.Math.Phys.}\ }\textbf {\bibinfo {volume} {1}},\ \bibinfo {pages} {429}
  (\bibinfo {year} {1960})}\BibitemShut {NoStop}%
\bibitem [{\citenamefont {Goncharov}()}]{GoncharovMixedTate}%
  \BibitemOpen
  \bibfield  {author} {\bibinfo {author} {\bibfnamefont {A.~B.}\ \bibnamefont
  {Goncharov}},\ }\href@noop {} {\ }\Eprint {http://arxiv.org/abs/math/0103059}
  {arXiv:math/0103059} \BibitemShut {NoStop}%
\bibitem [{\citenamefont {Goncharov}(2005)}]{Goncharov:2005sla}%
  \BibitemOpen
  \bibfield  {author} {\bibinfo {author} {\bibfnamefont {A.~B.}\ \bibnamefont
  {Goncharov}},\ }\href {\doibase 10.1215/S0012-7094-04-12822-2} {\bibfield
  {journal} {\bibinfo  {journal} {Duke Math.J.}\ }\textbf {\bibinfo {volume}
  {128}},\ \bibinfo {pages} {209} (\bibinfo {year} {2005})},\ \Eprint
  {http://arxiv.org/abs/math/0208144} {arXiv:math/0208144 [math.AG]}
  \BibitemShut {NoStop}%
\bibitem [{\citenamefont {Brown}(2009{\natexlab{a}})}]{Brown:2008um}%
  \BibitemOpen
  \bibfield  {author} {\bibinfo {author} {\bibfnamefont {F.}~\bibnamefont
  {Brown}},\ }\href {\doibase 10.1007/s00220-009-0740-5} {\bibfield  {journal}
  {\bibinfo  {journal} {Commun. Math. Phys.}\ }\textbf {\bibinfo {volume}
  {287}},\ \bibinfo {pages} {925} (\bibinfo {year} {2009}{\natexlab{a}})},\
  \Eprint {http://arxiv.org/abs/0804.1660} {arXiv:0804.1660 [math.AG]}
  \BibitemShut {NoStop}%
\bibitem [{\citenamefont {Anastasiou}\ \emph {et~al.}(2013)\citenamefont
  {Anastasiou}, \citenamefont {Duhr}, \citenamefont {Dulat},\ and\
  \citenamefont {Mistlberger}}]{Anastasiou:2013srw}%
  \BibitemOpen
  \bibfield  {author} {\bibinfo {author} {\bibfnamefont {C.}~\bibnamefont
  {Anastasiou}}, \bibinfo {author} {\bibfnamefont {C.}~\bibnamefont {Duhr}},
  \bibinfo {author} {\bibfnamefont {F.}~\bibnamefont {Dulat}}, \ and\ \bibinfo
  {author} {\bibfnamefont {B.}~\bibnamefont {Mistlberger}},\ }\href {\doibase
  10.1007/JHEP07(2013)003} {\bibfield  {journal} {\bibinfo  {journal} {JHEP}\
  }\textbf {\bibinfo {volume} {07}},\ \bibinfo {pages} {003} (\bibinfo {year}
  {2013})},\ \Eprint {http://arxiv.org/abs/1302.4379} {arXiv:1302.4379
  [hep-ph]} \BibitemShut {NoStop}%
\bibitem [{\citenamefont {Panzer}(2015)}]{Panzer:2014caa}%
  \BibitemOpen
  \bibfield  {author} {\bibinfo {author} {\bibfnamefont {E.}~\bibnamefont
  {Panzer}},\ }\href {\doibase 10.1016/j.cpc.2014.10.019} {\bibfield  {journal}
  {\bibinfo  {journal} {Comput. Phys. Commun.}\ }\textbf {\bibinfo {volume}
  {188}},\ \bibinfo {pages} {148} (\bibinfo {year} {2015})},\ \Eprint
  {http://arxiv.org/abs/1403.3385} {arXiv:1403.3385 [hep-th]} \BibitemShut
  {NoStop}%
\bibitem [{\citenamefont {Bogner}\ and\ \citenamefont
  {Brown}(2015)}]{Bogner:2014mha}%
  \BibitemOpen
  \bibfield  {author} {\bibinfo {author} {\bibfnamefont {C.}~\bibnamefont
  {Bogner}}\ and\ \bibinfo {author} {\bibfnamefont {F.}~\bibnamefont {Brown}},\
  }\href {\doibase 10.4310/CNTP.2015.v9.n1.a3} {\bibfield  {journal} {\bibinfo
  {journal} {Commun. Num. Theor. Phys.}\ }\textbf {\bibinfo {volume} {09}},\
  \bibinfo {pages} {189} (\bibinfo {year} {2015})},\ \Eprint
  {http://arxiv.org/abs/1408.1862} {arXiv:1408.1862 [hep-th]} \BibitemShut
  {NoStop}%
\bibitem [{\citenamefont {Bogner}(2016)}]{Bogner:2015nda}%
  \BibitemOpen
  \bibfield  {author} {\bibinfo {author} {\bibfnamefont {C.}~\bibnamefont
  {Bogner}},\ }\href {\doibase 10.1016/j.cpc.2016.02.033} {\bibfield  {journal}
  {\bibinfo  {journal} {Comput. Phys. Commun.}\ }\textbf {\bibinfo {volume}
  {203}},\ \bibinfo {pages} {339} (\bibinfo {year} {2016})},\ \Eprint
  {http://arxiv.org/abs/1510.04562} {arXiv:1510.04562 [physics.comp-ph]}
  \BibitemShut {NoStop}%
\bibitem [{\citenamefont {Ablinger}\ \emph {et~al.}(2014)\citenamefont
  {Ablinger}, \citenamefont {Bl{\"u}mlein}, \citenamefont {Raab}, \citenamefont
  {Schneider},\ and\ \citenamefont {Wi{\ss}brock}}]{Ablinger:2014yaa}%
  \BibitemOpen
  \bibfield  {author} {\bibinfo {author} {\bibfnamefont {J.}~\bibnamefont
  {Ablinger}}, \bibinfo {author} {\bibfnamefont {J.}~\bibnamefont
  {Bl{\"u}mlein}}, \bibinfo {author} {\bibfnamefont {C.}~\bibnamefont {Raab}},
  \bibinfo {author} {\bibfnamefont {C.}~\bibnamefont {Schneider}}, \ and\
  \bibinfo {author} {\bibfnamefont {F.}~\bibnamefont {Wi{\ss}brock}},\ }\href
  {\doibase 10.1016/j.nuclphysb.2014.04.007} {\bibfield  {journal} {\bibinfo
  {journal} {Nucl. Phys.}\ }\textbf {\bibinfo {volume} {B885}},\ \bibinfo
  {pages} {409} (\bibinfo {year} {2014})},\ \Eprint
  {http://arxiv.org/abs/1403.1137} {arXiv:1403.1137 [hep-ph]} \BibitemShut
  {NoStop}%
\bibitem [{\citenamefont {Henn}(2013)}]{Henn:2013pwa}%
  \BibitemOpen
  \bibfield  {author} {\bibinfo {author} {\bibfnamefont {J.~M.}\ \bibnamefont
  {Henn}},\ }\href {\doibase 10.1103/PhysRevLett.110.251601} {\bibfield
  {journal} {\bibinfo  {journal} {Phys.Rev.Lett.}\ }\textbf {\bibinfo {volume}
  {110}},\ \bibinfo {pages} {251601} (\bibinfo {year} {2013})},\ \Eprint
  {http://arxiv.org/abs/1304.1806} {arXiv:1304.1806 [hep-th]} \BibitemShut
  {NoStop}%
\bibitem [{\citenamefont {Goncharov}\ \emph {et~al.}(2010)\citenamefont
  {Goncharov}, \citenamefont {Spradlin}, \citenamefont {Vergu},\ and\
  \citenamefont {Volovich}}]{Goncharov:2010jf}%
  \BibitemOpen
  \bibfield  {author} {\bibinfo {author} {\bibfnamefont {A.~B.}\ \bibnamefont
  {Goncharov}}, \bibinfo {author} {\bibfnamefont {M.}~\bibnamefont {Spradlin}},
  \bibinfo {author} {\bibfnamefont {C.}~\bibnamefont {Vergu}}, \ and\ \bibinfo
  {author} {\bibfnamefont {A.}~\bibnamefont {Volovich}},\ }\href {\doibase
  10.1103/PhysRevLett.105.151605} {\bibfield  {journal} {\bibinfo  {journal}
  {Phys.Rev.Lett.}\ }\textbf {\bibinfo {volume} {105}},\ \bibinfo {pages}
  {151605} (\bibinfo {year} {2010})},\ \Eprint {http://arxiv.org/abs/1006.5703}
  {arXiv:1006.5703 [hep-th]} \BibitemShut {NoStop}%
\bibitem [{\citenamefont {Remiddi}\ and\ \citenamefont
  {Vermaseren}(2000)}]{Remiddi:1999ew}%
  \BibitemOpen
  \bibfield  {author} {\bibinfo {author} {\bibfnamefont {E.}~\bibnamefont
  {Remiddi}}\ and\ \bibinfo {author} {\bibfnamefont {J.}~\bibnamefont
  {Vermaseren}},\ }\href {\doibase 10.1142/S0217751X00000367} {\bibfield
  {journal} {\bibinfo  {journal} {Int.J.Mod.Phys.}\ }\textbf {\bibinfo {volume}
  {A15}},\ \bibinfo {pages} {725} (\bibinfo {year} {2000})},\ \Eprint
  {http://arxiv.org/abs/hep-ph/9905237} {arXiv:hep-ph/9905237 [hep-ph]}
  \BibitemShut {NoStop}%
\bibitem [{\citenamefont {Brown}(2009{\natexlab{b}})}]{Brown:2009qja}%
  \BibitemOpen
  \bibfield  {author} {\bibinfo {author} {\bibfnamefont {F.~C.}\ \bibnamefont
  {Brown}},\ }\href@noop {} {\bibfield  {journal} {\bibinfo  {journal} {Annales
  Sci.Ecole Norm.Sup.}\ }\textbf {\bibinfo {volume} {42}},\ \bibinfo {pages}
  {371} (\bibinfo {year} {2009}{\natexlab{b}})},\ \Eprint
  {http://arxiv.org/abs/math/0606419} {arXiv:math/0606419 [math.AG]}
  \BibitemShut {NoStop}%
\bibitem [{\citenamefont {Brown}(2011)}]{Brown:2011ik}%
  \BibitemOpen
  \bibfield  {author} {\bibinfo {author} {\bibfnamefont {F.}~\bibnamefont
  {Brown}},\ }\href@noop {} {\  (\bibinfo {year} {2011})},\ \Eprint
  {http://arxiv.org/abs/1102.1310} {arXiv:1102.1310 [math.NT]} \BibitemShut
  {NoStop}%
\bibitem [{\citenamefont {Duhr}\ \emph {et~al.}(2012)\citenamefont {Duhr},
  \citenamefont {Gangl},\ and\ \citenamefont {Rhodes}}]{Duhr:2011zq}%
  \BibitemOpen
  \bibfield  {author} {\bibinfo {author} {\bibfnamefont {C.}~\bibnamefont
  {Duhr}}, \bibinfo {author} {\bibfnamefont {H.}~\bibnamefont {Gangl}}, \ and\
  \bibinfo {author} {\bibfnamefont {J.~R.}\ \bibnamefont {Rhodes}},\ }\href
  {\doibase 10.1007/JHEP10(2012)075} {\bibfield  {journal} {\bibinfo  {journal}
  {JHEP}\ }\textbf {\bibinfo {volume} {1210}},\ \bibinfo {pages} {075}
  (\bibinfo {year} {2012})},\ \Eprint {http://arxiv.org/abs/1110.0458}
  {arXiv:1110.0458 [math-ph]} \BibitemShut {NoStop}%
\bibitem [{\citenamefont {Duhr}(2012)}]{Duhr:2012fh}%
  \BibitemOpen
  \bibfield  {author} {\bibinfo {author} {\bibfnamefont {C.}~\bibnamefont
  {Duhr}},\ }\href {\doibase 10.1007/JHEP08(2012)043} {\bibfield  {journal}
  {\bibinfo  {journal} {JHEP}\ }\textbf {\bibinfo {volume} {1208}},\ \bibinfo
  {pages} {043} (\bibinfo {year} {2012})},\ \Eprint
  {http://arxiv.org/abs/1203.0454} {arXiv:1203.0454 [hep-ph]} \BibitemShut
  {NoStop}%
\bibitem [{\citenamefont {Caffo}\ \emph {et~al.}(1998)\citenamefont {Caffo},
  \citenamefont {Czyz}, \citenamefont {Laporta},\ and\ \citenamefont
  {Remiddi}}]{Caffo:1998du}%
  \BibitemOpen
  \bibfield  {author} {\bibinfo {author} {\bibfnamefont {M.}~\bibnamefont
  {Caffo}}, \bibinfo {author} {\bibfnamefont {H.}~\bibnamefont {Czyz}},
  \bibinfo {author} {\bibfnamefont {S.}~\bibnamefont {Laporta}}, \ and\
  \bibinfo {author} {\bibfnamefont {E.}~\bibnamefont {Remiddi}},\ }\href@noop
  {} {\bibfield  {journal} {\bibinfo  {journal} {Nuovo Cim.}\ }\textbf
  {\bibinfo {volume} {A111}},\ \bibinfo {pages} {365} (\bibinfo {year}
  {1998})},\ \Eprint {http://arxiv.org/abs/hep-th/9805118}
  {arXiv:hep-th/9805118 [hep-th]} \BibitemShut {NoStop}%
\bibitem [{\citenamefont {Bloch}\ and\ \citenamefont
  {Vanhove}(2015)}]{Bloch:2013tra}%
  \BibitemOpen
  \bibfield  {author} {\bibinfo {author} {\bibfnamefont {S.}~\bibnamefont
  {Bloch}}\ and\ \bibinfo {author} {\bibfnamefont {P.}~\bibnamefont
  {Vanhove}},\ }\href {\doibase 10.1016/j.jnt.2014.09.032} {\bibfield
  {journal} {\bibinfo  {journal} {J. Number Theor.}\ }\textbf {\bibinfo
  {volume} {148}},\ \bibinfo {pages} {328} (\bibinfo {year} {2015})},\ \Eprint
  {http://arxiv.org/abs/1309.5865} {arXiv:1309.5865 [hep-th]} \BibitemShut
  {NoStop}%
\bibitem [{\citenamefont {Adams}\ \emph {et~al.}(2013)\citenamefont {Adams},
  \citenamefont {Bogner},\ and\ \citenamefont {Weinzierl}}]{Adams:2013kgc}%
  \BibitemOpen
  \bibfield  {author} {\bibinfo {author} {\bibfnamefont {L.}~\bibnamefont
  {Adams}}, \bibinfo {author} {\bibfnamefont {C.}~\bibnamefont {Bogner}}, \
  and\ \bibinfo {author} {\bibfnamefont {S.}~\bibnamefont {Weinzierl}},\ }\href
  {\doibase 10.1063/1.4804996} {\bibfield  {journal} {\bibinfo  {journal} {J.
  Math. Phys.}\ }\textbf {\bibinfo {volume} {54}},\ \bibinfo {pages} {052303}
  (\bibinfo {year} {2013})},\ \Eprint {http://arxiv.org/abs/1302.7004}
  {arXiv:1302.7004 [hep-ph]} \BibitemShut {NoStop}%
\bibitem [{\citenamefont {Bloch}\ \emph {et~al.}(2015)\citenamefont {Bloch},
  \citenamefont {Kerr},\ and\ \citenamefont {Vanhove}}]{Bloch:2014qca}%
  \BibitemOpen
  \bibfield  {author} {\bibinfo {author} {\bibfnamefont {S.}~\bibnamefont
  {Bloch}}, \bibinfo {author} {\bibfnamefont {M.}~\bibnamefont {Kerr}}, \ and\
  \bibinfo {author} {\bibfnamefont {P.}~\bibnamefont {Vanhove}},\ }\href
  {\doibase 10.1112/S0010437X15007472} {\bibfield  {journal} {\bibinfo
  {journal} {Compos. Math.}\ }\textbf {\bibinfo {volume} {151}},\ \bibinfo
  {pages} {2329} (\bibinfo {year} {2015})},\ \Eprint
  {http://arxiv.org/abs/1406.2664} {arXiv:1406.2664 [hep-th]} \BibitemShut
  {NoStop}%
\bibitem [{\citenamefont {Adams}\ \emph {et~al.}(2014)\citenamefont {Adams},
  \citenamefont {Bogner},\ and\ \citenamefont {Weinzierl}}]{Adams:2014vja}%
  \BibitemOpen
  \bibfield  {author} {\bibinfo {author} {\bibfnamefont {L.}~\bibnamefont
  {Adams}}, \bibinfo {author} {\bibfnamefont {C.}~\bibnamefont {Bogner}}, \
  and\ \bibinfo {author} {\bibfnamefont {S.}~\bibnamefont {Weinzierl}},\ }\href
  {\doibase 10.1063/1.4896563} {\bibfield  {journal} {\bibinfo  {journal}
  {J.Math.Phys.}\ }\textbf {\bibinfo {volume} {55}},\ \bibinfo {pages} {102301}
  (\bibinfo {year} {2014})},\ \Eprint {http://arxiv.org/abs/1405.5640}
  {arXiv:1405.5640 [hep-ph]} \BibitemShut {NoStop}%
\bibitem [{\citenamefont {Adams}\ \emph {et~al.}(2015)\citenamefont {Adams},
  \citenamefont {Bogner},\ and\ \citenamefont {Weinzierl}}]{Adams:2015gva}%
  \BibitemOpen
  \bibfield  {author} {\bibinfo {author} {\bibfnamefont {L.}~\bibnamefont
  {Adams}}, \bibinfo {author} {\bibfnamefont {C.}~\bibnamefont {Bogner}}, \
  and\ \bibinfo {author} {\bibfnamefont {S.}~\bibnamefont {Weinzierl}},\ }\href
  {\doibase 10.1063/1.4926985} {\bibfield  {journal} {\bibinfo  {journal} {J.
  Math. Phys.}\ }\textbf {\bibinfo {volume} {56}},\ \bibinfo {pages} {072303}
  (\bibinfo {year} {2015})},\ \Eprint {http://arxiv.org/abs/1504.03255}
  {arXiv:1504.03255 [hep-ph]} \BibitemShut {NoStop}%
\bibitem [{\citenamefont {Adams}\ \emph
  {et~al.}(2016{\natexlab{a}})\citenamefont {Adams}, \citenamefont {Bogner},\
  and\ \citenamefont {Weinzierl}}]{Adams:2015ydq}%
  \BibitemOpen
  \bibfield  {author} {\bibinfo {author} {\bibfnamefont {L.}~\bibnamefont
  {Adams}}, \bibinfo {author} {\bibfnamefont {C.}~\bibnamefont {Bogner}}, \
  and\ \bibinfo {author} {\bibfnamefont {S.}~\bibnamefont {Weinzierl}},\ }\href
  {\doibase 10.1063/1.4944722} {\bibfield  {journal} {\bibinfo  {journal} {J.
  Math. Phys.}\ }\textbf {\bibinfo {volume} {57}},\ \bibinfo {pages} {032304}
  (\bibinfo {year} {2016}{\natexlab{a}})},\ \Eprint
  {http://arxiv.org/abs/1512.05630} {arXiv:1512.05630 [hep-ph]} \BibitemShut
  {NoStop}%
\bibitem [{\citenamefont {Bloch}\ \emph {et~al.}(2016)\citenamefont {Bloch},
  \citenamefont {Kerr},\ and\ \citenamefont {Vanhove}}]{Bloch:2016izu}%
  \BibitemOpen
  \bibfield  {author} {\bibinfo {author} {\bibfnamefont {S.}~\bibnamefont
  {Bloch}}, \bibinfo {author} {\bibfnamefont {M.}~\bibnamefont {Kerr}}, \ and\
  \bibinfo {author} {\bibfnamefont {P.}~\bibnamefont {Vanhove}},\ }\href@noop
  {} {\  (\bibinfo {year} {2016})},\ \Eprint {http://arxiv.org/abs/1601.08181}
  {arXiv:1601.08181 [hep-th]} \BibitemShut {NoStop}%
\bibitem [{\citenamefont {Adams}\ \emph
  {et~al.}(2016{\natexlab{b}})\citenamefont {Adams}, \citenamefont {Bogner},
  \citenamefont {Schweitzer},\ and\ \citenamefont {Weinzierl}}]{Adams:2016xah}%
  \BibitemOpen
  \bibfield  {author} {\bibinfo {author} {\bibfnamefont {L.}~\bibnamefont
  {Adams}}, \bibinfo {author} {\bibfnamefont {C.}~\bibnamefont {Bogner}},
  \bibinfo {author} {\bibfnamefont {A.}~\bibnamefont {Schweitzer}}, \ and\
  \bibinfo {author} {\bibfnamefont {S.}~\bibnamefont {Weinzierl}},\ }\href
  {\doibase 10.1063/1.4969060} {\bibfield  {journal} {\bibinfo  {journal} {{J.
  Math. Phys.}}\ }\textbf {\bibinfo {volume} {57}},\ \bibinfo {pages} {122302}
  (\bibinfo {year} {2016}{\natexlab{b}})},\ \Eprint
  {http://arxiv.org/abs/1607.01571} {arXiv:1607.01571 [hep-ph]} \BibitemShut
  {NoStop}%
\bibitem [{\citenamefont {Remiddi}\ and\ \citenamefont
  {Tancredi}(2016)}]{Remiddi:2016gno}%
  \BibitemOpen
  \bibfield  {author} {\bibinfo {author} {\bibfnamefont {E.}~\bibnamefont
  {Remiddi}}\ and\ \bibinfo {author} {\bibfnamefont {L.}~\bibnamefont
  {Tancredi}},\ }\href {\doibase 10.1016/j.nuclphysb.2016.04.013} {\bibfield
  {journal} {\bibinfo  {journal} {Nucl. Phys.}\ }\textbf {\bibinfo {volume}
  {B907}},\ \bibinfo {pages} {400} (\bibinfo {year} {2016})},\ \Eprint
  {http://arxiv.org/abs/1602.01481} {arXiv:1602.01481 [hep-ph]} \BibitemShut
  {NoStop}%
\bibitem [{\citenamefont {Primo}\ and\ \citenamefont
  {Tancredi}(2017)}]{Primo:2016ebd}%
  \BibitemOpen
  \bibfield  {author} {\bibinfo {author} {\bibfnamefont {A.}~\bibnamefont
  {Primo}}\ and\ \bibinfo {author} {\bibfnamefont {L.}~\bibnamefont
  {Tancredi}},\ }\href {\doibase 10.1016/j.nuclphysb.2016.12.021} {\bibfield
  {journal} {\bibinfo  {journal} {Nucl. Phys.}\ }\textbf {\bibinfo {volume}
  {B916}},\ \bibinfo {pages} {94} (\bibinfo {year} {2017})},\ \Eprint
  {http://arxiv.org/abs/1610.08397} {arXiv:1610.08397 [hep-ph]} \BibitemShut
  {NoStop}%
\bibitem [{\citenamefont {von Manteuffel}\ and\ \citenamefont
  {Tancredi}(2017)}]{vonManteuffel:2017hms}%
  \BibitemOpen
  \bibfield  {author} {\bibinfo {author} {\bibfnamefont {A.}~\bibnamefont {von
  Manteuffel}}\ and\ \bibinfo {author} {\bibfnamefont {L.}~\bibnamefont
  {Tancredi}},\ }\href@noop {} {\  (\bibinfo {year} {2017})},\ \Eprint
  {http://arxiv.org/abs/1701.05905} {arXiv:1701.05905 [hep-ph]} \BibitemShut
  {NoStop}%
\bibitem [{\citenamefont {Brown}(2015{\natexlab{a}})}]{Brown:motivicperiods}%
  \BibitemOpen
  \bibfield  {author} {\bibinfo {author} {\bibfnamefont {F.}~\bibnamefont
  {Brown}},\ }\href@noop {} {\  (\bibinfo {year} {2015}{\natexlab{a}})},\
  \Eprint {http://arxiv.org/abs/1512.06410} {arXiv:1512.06410 [math.NT]}
  \BibitemShut {NoStop}%
\bibitem [{\citenamefont {Brown}()}]{BrownTalk}%
  \BibitemOpen
  \bibfield  {author} {\bibinfo {author} {\bibfnamefont {F.~C.}\ \bibnamefont
  {Brown}},\ }\href@noop {} {}\Eprint
  {http://arxiv.org/abs/http://www.ihes.fr/\textasciitilde{}brown/SMFBrown.pdf}
  {http://www.ihes.fr/\textasciitilde{}brown/SMFBrown.pdf} \BibitemShut
  {NoStop}%
\bibitem [{\citenamefont {Moch}\ \emph {et~al.}(2002)\citenamefont {Moch},
  \citenamefont {Uwer},\ and\ \citenamefont {Weinzierl}}]{Moch:2001zr}%
  \BibitemOpen
  \bibfield  {author} {\bibinfo {author} {\bibfnamefont {S.}~\bibnamefont
  {Moch}}, \bibinfo {author} {\bibfnamefont {P.}~\bibnamefont {Uwer}}, \ and\
  \bibinfo {author} {\bibfnamefont {S.}~\bibnamefont {Weinzierl}},\ }\href
  {\doibase 10.1063/1.1471366} {\bibfield  {journal} {\bibinfo  {journal} {J.
  Math. Phys.}\ }\textbf {\bibinfo {volume} {43}},\ \bibinfo {pages} {3363}
  (\bibinfo {year} {2002})},\ \Eprint {http://arxiv.org/abs/hep-ph/0110083}
  {arXiv:hep-ph/0110083 [hep-ph]} \BibitemShut {NoStop}%
\bibitem [{\citenamefont {Moch}\ and\ \citenamefont
  {Uwer}(2006)}]{Moch:2005uc}%
  \BibitemOpen
  \bibfield  {author} {\bibinfo {author} {\bibfnamefont {S.}~\bibnamefont
  {Moch}}\ and\ \bibinfo {author} {\bibfnamefont {P.}~\bibnamefont {Uwer}},\
  }\href {\doibase 10.1016/j.cpc.2005.12.014} {\bibfield  {journal} {\bibinfo
  {journal} {Comput. Phys. Commun.}\ }\textbf {\bibinfo {volume} {174}},\
  \bibinfo {pages} {759} (\bibinfo {year} {2006})},\ \Eprint
  {http://arxiv.org/abs/math-ph/0508008} {arXiv:math-ph/0508008 [math-ph]}
  \BibitemShut {NoStop}%
\bibitem [{\citenamefont {Huber}\ and\ \citenamefont
  {Maitre}(2006)}]{Huber:2005yg}%
  \BibitemOpen
  \bibfield  {author} {\bibinfo {author} {\bibfnamefont {T.}~\bibnamefont
  {Huber}}\ and\ \bibinfo {author} {\bibfnamefont {D.}~\bibnamefont {Maitre}},\
  }\href {\doibase 10.1016/j.cpc.2006.01.007} {\bibfield  {journal} {\bibinfo
  {journal} {Comput. Phys. Commun.}\ }\textbf {\bibinfo {volume} {175}},\
  \bibinfo {pages} {122} (\bibinfo {year} {2006})},\ \Eprint
  {http://arxiv.org/abs/hep-ph/0507094} {arXiv:hep-ph/0507094} \BibitemShut
  {NoStop}%
\bibitem [{\citenamefont {Huber}\ and\ \citenamefont
  {Maitre}(2008)}]{Huber:2007dx}%
  \BibitemOpen
  \bibfield  {author} {\bibinfo {author} {\bibfnamefont {T.}~\bibnamefont
  {Huber}}\ and\ \bibinfo {author} {\bibfnamefont {D.}~\bibnamefont {Maitre}},\
  }\href {\doibase 10.1016/j.cpc.2007.12.008} {\bibfield  {journal} {\bibinfo
  {journal} {Comput.Phys.Commun.}\ }\textbf {\bibinfo {volume} {178}},\
  \bibinfo {pages} {755} (\bibinfo {year} {2008})},\ \Eprint
  {http://arxiv.org/abs/0708.2443} {arXiv:0708.2443 [hep-ph]} \BibitemShut
  {NoStop}%
\bibitem [{\citenamefont {Tkachov}(1981)}]{Tkachov:1981wb}%
  \BibitemOpen
  \bibfield  {author} {\bibinfo {author} {\bibfnamefont {F.~V.}\ \bibnamefont
  {Tkachov}},\ }\href {\doibase 10.1016/0370-2693(81)90288-4} {\bibfield
  {journal} {\bibinfo  {journal} {Phys. Lett.}\ }\textbf {\bibinfo {volume}
  {B100}},\ \bibinfo {pages} {65} (\bibinfo {year} {1981})}\BibitemShut
  {NoStop}%
\bibitem [{\citenamefont {Chetyrkin}\ and\ \citenamefont
  {Tkachov}(1981)}]{Chetyrkin:1981qh}%
  \BibitemOpen
  \bibfield  {author} {\bibinfo {author} {\bibfnamefont {K.}~\bibnamefont
  {Chetyrkin}}\ and\ \bibinfo {author} {\bibfnamefont {F.}~\bibnamefont
  {Tkachov}},\ }\href {\doibase 10.1016/0550-3213(81)90199-1} {\bibfield
  {journal} {\bibinfo  {journal} {Nucl.Phys.}\ }\textbf {\bibinfo {volume}
  {B192}},\ \bibinfo {pages} {159} (\bibinfo {year} {1981})}\BibitemShut
  {NoStop}%
\bibitem [{\citenamefont {Bern}\ \emph {et~al.}(1993)\citenamefont {Bern},
  \citenamefont {Dixon},\ and\ \citenamefont {Kosower}}]{Bern:1992em}%
  \BibitemOpen
  \bibfield  {author} {\bibinfo {author} {\bibfnamefont {Z.}~\bibnamefont
  {Bern}}, \bibinfo {author} {\bibfnamefont {L.~J.}\ \bibnamefont {Dixon}}, \
  and\ \bibinfo {author} {\bibfnamefont {D.~A.}\ \bibnamefont {Kosower}},\
  }\href {\doibase 10.1016/0370-2693(93)90469-X, 10.1016/0370-2693(93)90400-C}
  {\bibfield  {journal} {\bibinfo  {journal} {Phys. Lett.}\ }\textbf {\bibinfo
  {volume} {B302}},\ \bibinfo {pages} {299} (\bibinfo {year} {1993})},\
  \bibinfo {note} {[Erratum: Phys. Lett.B318,649(1993)]},\ \Eprint
  {http://arxiv.org/abs/hep-ph/9212308} {arXiv:hep-ph/9212308 [hep-ph]}
  \BibitemShut {NoStop}%
\bibitem [{\citenamefont {Tarasov}(1996)}]{Tarasov:1996br}%
  \BibitemOpen
  \bibfield  {author} {\bibinfo {author} {\bibfnamefont {O.~V.}\ \bibnamefont
  {Tarasov}},\ }\href {\doibase 10.1103/PhysRevD.54.6479} {\bibfield  {journal}
  {\bibinfo  {journal} {Phys. Rev.}\ }\textbf {\bibinfo {volume} {D54}},\
  \bibinfo {pages} {6479} (\bibinfo {year} {1996})},\ \Eprint
  {http://arxiv.org/abs/hep-th/9606018} {arXiv:hep-th/9606018 [hep-th]}
  \BibitemShut {NoStop}%
\bibitem [{\citenamefont {Lee}(2010)}]{Lee:2009dh}%
  \BibitemOpen
  \bibfield  {author} {\bibinfo {author} {\bibfnamefont {R.~N.}\ \bibnamefont
  {Lee}},\ }\href {\doibase 10.1016/j.nuclphysb.2009.12.025} {\bibfield
  {journal} {\bibinfo  {journal} {Nucl. Phys.}\ }\textbf {\bibinfo {volume}
  {B830}},\ \bibinfo {pages} {474} (\bibinfo {year} {2010})},\ \Eprint
  {http://arxiv.org/abs/0911.0252} {arXiv:0911.0252 [hep-ph]} \BibitemShut
  {NoStop}%
\bibitem [{\citenamefont {Abreu}\ \emph {et~al.}(2017)\citenamefont {Abreu},
  \citenamefont {Britto}, \citenamefont {Duhr},\ and\ \citenamefont
  {Gardi}}]{Abreu:2017ptx}%
  \BibitemOpen
  \bibfield  {author} {\bibinfo {author} {\bibfnamefont {S.}~\bibnamefont
  {Abreu}}, \bibinfo {author} {\bibfnamefont {R.}~\bibnamefont {Britto}},
  \bibinfo {author} {\bibfnamefont {C.}~\bibnamefont {Duhr}}, \ and\ \bibinfo
  {author} {\bibfnamefont {E.}~\bibnamefont {Gardi}},\ }\href@noop {} {\
  (\bibinfo {year} {2017})},\ \Eprint {http://arxiv.org/abs/1702.03163}
  {arXiv:1702.03163 [hep-th]} \BibitemShut {NoStop}%
\bibitem [{\citenamefont {Fotiadi}\ \emph {et~al.}(1965)\citenamefont
  {Fotiadi}, \citenamefont {Froissart}, \citenamefont {Lascoux},\ and\
  \citenamefont {Pham}}]{PhamCompact}%
  \BibitemOpen
  \bibfield  {author} {\bibinfo {author} {\bibfnamefont {D.}~\bibnamefont
  {Fotiadi}}, \bibinfo {author} {\bibfnamefont {M.}~\bibnamefont {Froissart}},
  \bibinfo {author} {\bibfnamefont {J.}~\bibnamefont {Lascoux}}, \ and\
  \bibinfo {author} {\bibfnamefont {F.}~\bibnamefont {Pham}},\ }\href@noop {}
  {\bibfield  {journal} {\bibinfo  {journal} {Topology}\ }\textbf {\bibinfo
  {volume} {4}},\ \bibinfo {pages} {159} (\bibinfo {year} {1965})}\BibitemShut
  {NoStop}%
\bibitem [{\citenamefont {Fotiadi}\ and\ \citenamefont
  {Pham}(1966)}]{Froissart}%
  \BibitemOpen
  \bibfield  {author} {\bibinfo {author} {\bibfnamefont {D.}~\bibnamefont
  {Fotiadi}}\ and\ \bibinfo {author} {\bibfnamefont {F.}~\bibnamefont {Pham}},\
  }in\ \href@noop {} {\emph {\bibinfo {booktitle} {{Homology and Feynman
  integrals}}}},\ \bibinfo {editor} {edited by\ \bibinfo {editor}
  {\bibfnamefont {R.~C.}\ \bibnamefont {Hwa}}\ and\ \bibinfo {editor}
  {\bibfnamefont {V.~L.}\ \bibnamefont {Teplitz}}}\ (\bibinfo  {publisher}
  {W.~A.~Benjamin Inc.},\ \bibinfo {year} {1966})\BibitemShut {NoStop}%
\bibitem [{\citenamefont {Hwa}\ and\ \citenamefont {Teplitz}(1966)}]{Teplitz}%
  \BibitemOpen
  \bibfield  {author} {\bibinfo {author} {\bibfnamefont {R.~C.}\ \bibnamefont
  {Hwa}}\ and\ \bibinfo {author} {\bibfnamefont {V.~L.}\ \bibnamefont
  {Teplitz}},\ }\href@noop {} {\emph {\bibinfo {title} {{Homology and Feynman
  integrals}}}}\ (\bibinfo  {publisher} {W.~A.~Benjamin Inc.},\ \bibinfo {year}
  {1966})\BibitemShut {NoStop}%
\bibitem [{\citenamefont {Abreu}\ \emph {et~al.}(pear)\citenamefont {Abreu},
  \citenamefont {Britto}, \citenamefont {Duhr},\ and\ \citenamefont
  {Gardi}}]{LongDiag}%
  \BibitemOpen
  \bibfield  {author} {\bibinfo {author} {\bibfnamefont {S.}~\bibnamefont
  {Abreu}}, \bibinfo {author} {\bibfnamefont {R.}~\bibnamefont {Britto}},
  \bibinfo {author} {\bibfnamefont {C.}~\bibnamefont {Duhr}}, \ and\ \bibinfo
  {author} {\bibfnamefont {E.}~\bibnamefont {Gardi}},\ }\href@noop {} {\
  (\bibinfo {year} {To appear})}\BibitemShut {NoStop}%
\bibitem [{\citenamefont {Gaiotto}\ \emph {et~al.}(2011)\citenamefont
  {Gaiotto}, \citenamefont {Maldacena}, \citenamefont {Sever},\ and\
  \citenamefont {Vieira}}]{Gaiotto:2011dt}%
  \BibitemOpen
  \bibfield  {author} {\bibinfo {author} {\bibfnamefont {D.}~\bibnamefont
  {Gaiotto}}, \bibinfo {author} {\bibfnamefont {J.}~\bibnamefont {Maldacena}},
  \bibinfo {author} {\bibfnamefont {A.}~\bibnamefont {Sever}}, \ and\ \bibinfo
  {author} {\bibfnamefont {P.}~\bibnamefont {Vieira}},\ }\href {\doibase
  10.1007/JHEP12(2011)011} {\bibfield  {journal} {\bibinfo  {journal} {JHEP}\
  }\textbf {\bibinfo {volume} {1112}},\ \bibinfo {pages} {011} (\bibinfo {year}
  {2011})},\ \Eprint {http://arxiv.org/abs/1102.0062} {arXiv:1102.0062
  [hep-th]} \BibitemShut {NoStop}%
\bibitem [{\citenamefont {Abreu}\ \emph {et~al.}(2015)\citenamefont {Abreu},
  \citenamefont {Britto},\ and\ \citenamefont {Gr{\"o}nqvist}}]{Abreu:2015zaa}%
  \BibitemOpen
  \bibfield  {author} {\bibinfo {author} {\bibfnamefont {S.}~\bibnamefont
  {Abreu}}, \bibinfo {author} {\bibfnamefont {R.}~\bibnamefont {Britto}}, \
  and\ \bibinfo {author} {\bibfnamefont {H.}~\bibnamefont {Gr{\"o}nqvist}},\
  }\href {\doibase 10.1007/JHEP07(2015)111} {\bibfield  {journal} {\bibinfo
  {journal} {JHEP}\ }\textbf {\bibinfo {volume} {07}},\ \bibinfo {pages} {111}
  (\bibinfo {year} {2015})},\ \Eprint {http://arxiv.org/abs/1504.00206}
  {arXiv:1504.00206 [hep-th]} \BibitemShut {NoStop}%
\bibitem [{\citenamefont {Brown}(2015{\natexlab{b}})}]{Brown:2015fyf}%
  \BibitemOpen
  \bibfield  {author} {\bibinfo {author} {\bibfnamefont {F.}~\bibnamefont
  {Brown}},\ }\href@noop {} {\  (\bibinfo {year} {2015}{\natexlab{b}})},\
  \Eprint {http://arxiv.org/abs/1512.06409} {arXiv:1512.06409 [math-ph]}
  \BibitemShut {NoStop}%
\bibitem [{\citenamefont {Steinmann}(1960{\natexlab{a}})}]{steinmann1}%
  \BibitemOpen
  \bibfield  {author} {\bibinfo {author} {\bibfnamefont {O.}~\bibnamefont
  {Steinmann}},\ }\href@noop {} {\bibfield  {journal} {\bibinfo  {journal}
  {Helv.Phys.Acta.}\ }\textbf {\bibinfo {volume} {33}},\ \bibinfo {pages} {257}
  (\bibinfo {year} {1960}{\natexlab{a}})}\BibitemShut {NoStop}%
\bibitem [{\citenamefont {Steinmann}(1960{\natexlab{b}})}]{steinmann2}%
  \BibitemOpen
  \bibfield  {author} {\bibinfo {author} {\bibfnamefont {O.}~\bibnamefont
  {Steinmann}},\ }\href@noop {} {\bibfield  {journal} {\bibinfo  {journal}
  {Helv.Phys.Acta.}\ }\textbf {\bibinfo {volume} {33}},\ \bibinfo {pages} {347}
  (\bibinfo {year} {1960}{\natexlab{b}})}\BibitemShut {NoStop}%
\bibitem [{\citenamefont {Cahill}\ and\ \citenamefont
  {Stapp}(1975)}]{Cahill:1973qp}%
  \BibitemOpen
  \bibfield  {author} {\bibinfo {author} {\bibfnamefont {K.~E.}\ \bibnamefont
  {Cahill}}\ and\ \bibinfo {author} {\bibfnamefont {H.~P.}\ \bibnamefont
  {Stapp}},\ }\href {\doibase 10.1016/0003-4916(75)90006-8} {\bibfield
  {journal} {\bibinfo  {journal} {Annals Phys.}\ }\textbf {\bibinfo {volume}
  {90}},\ \bibinfo {pages} {438} (\bibinfo {year} {1975})}\BibitemShut
  {NoStop}%
\bibitem [{\citenamefont {Caron-Huot}\ \emph {et~al.}(2016)\citenamefont
  {Caron-Huot}, \citenamefont {Dixon}, \citenamefont {McLeod},\ and\
  \citenamefont {von Hippel}}]{Caron-Huot:2016owq}%
  \BibitemOpen
  \bibfield  {author} {\bibinfo {author} {\bibfnamefont {S.}~\bibnamefont
  {Caron-Huot}}, \bibinfo {author} {\bibfnamefont {L.~J.}\ \bibnamefont
  {Dixon}}, \bibinfo {author} {\bibfnamefont {A.}~\bibnamefont {McLeod}}, \
  and\ \bibinfo {author} {\bibfnamefont {M.}~\bibnamefont {von Hippel}},\
  }\href {\doibase 10.1103/PhysRevLett.117.241601} {\bibfield  {journal}
  {\bibinfo  {journal} {Phys. Rev. Lett.}\ }\textbf {\bibinfo {volume} {117}},\
  \bibinfo {pages} {241601} (\bibinfo {year} {2016})},\ \Eprint
  {http://arxiv.org/abs/1609.00669} {arXiv:1609.00669 [hep-th]} \BibitemShut
  {NoStop}%
\bibitem [{\citenamefont {Abreu}\ \emph {et~al.}(2014)\citenamefont {Abreu},
  \citenamefont {Britto}, \citenamefont {Duhr},\ and\ \citenamefont
  {Gardi}}]{Abreu:2014cla}%
  \BibitemOpen
  \bibfield  {author} {\bibinfo {author} {\bibfnamefont {S.}~\bibnamefont
  {Abreu}}, \bibinfo {author} {\bibfnamefont {R.}~\bibnamefont {Britto}},
  \bibinfo {author} {\bibfnamefont {C.}~\bibnamefont {Duhr}}, \ and\ \bibinfo
  {author} {\bibfnamefont {E.}~\bibnamefont {Gardi}},\ }\href {\doibase
  10.1007/JHEP10(2014)125} {\bibfield  {journal} {\bibinfo  {journal} {JHEP}\
  }\textbf {\bibinfo {volume} {1410}},\ \bibinfo {pages} {125} (\bibinfo {year}
  {2014})},\ \Eprint {http://arxiv.org/abs/1401.3546} {arXiv:1401.3546
  [hep-th]} \BibitemShut {NoStop}%
\bibitem [{\citenamefont {Caron-Huot}\ and\ \citenamefont
  {Larsen}(2012)}]{CaronHuot:2012ab}%
  \BibitemOpen
  \bibfield  {author} {\bibinfo {author} {\bibfnamefont {S.}~\bibnamefont
  {Caron-Huot}}\ and\ \bibinfo {author} {\bibfnamefont {K.~J.}\ \bibnamefont
  {Larsen}},\ }\href {\doibase 10.1007/JHEP10(2012)026} {\bibfield  {journal}
  {\bibinfo  {journal} {JHEP}\ }\textbf {\bibinfo {volume} {1210}},\ \bibinfo
  {pages} {026} (\bibinfo {year} {2012})},\ \Eprint
  {http://arxiv.org/abs/1205.0801} {arXiv:1205.0801 [hep-ph]} \BibitemShut
  {NoStop}%
\bibitem [{\citenamefont {Federbush}(1965)}]{Federbush}%
  \BibitemOpen
  \bibfield  {author} {\bibinfo {author} {\bibfnamefont {P.}~\bibnamefont
  {Federbush}},\ }\href@noop {} {\bibfield  {journal} {\bibinfo  {journal} {{J.
  Math. Phys.}}\ }\textbf {\bibinfo {volume} {6}},\ \bibinfo {pages} {941}
  (\bibinfo {year} {1965})}\BibitemShut {NoStop}%
\bibitem [{\citenamefont {Bloch}\ \emph {et~al.}(2006)\citenamefont {Bloch},
  \citenamefont {Esnault},\ and\ \citenamefont {Kreimer}}]{Bloch:2005bh}%
  \BibitemOpen
  \bibfield  {author} {\bibinfo {author} {\bibfnamefont {S.}~\bibnamefont
  {Bloch}}, \bibinfo {author} {\bibfnamefont {H.}~\bibnamefont {Esnault}}, \
  and\ \bibinfo {author} {\bibfnamefont {D.}~\bibnamefont {Kreimer}},\ }\href
  {\doibase 10.1007/s00220-006-0040-2} {\bibfield  {journal} {\bibinfo
  {journal} {Commun. Math. Phys.}\ }\textbf {\bibinfo {volume} {267}},\
  \bibinfo {pages} {181} (\bibinfo {year} {2006})},\ \Eprint
  {http://arxiv.org/abs/math/0510011} {arXiv:math/0510011 [math-ag]}
  \BibitemShut {NoStop}%
\bibitem [{\citenamefont {Kreimer}(1998)}]{Kreimer:1997dp}%
  \BibitemOpen
  \bibfield  {author} {\bibinfo {author} {\bibfnamefont {D.}~\bibnamefont
  {Kreimer}},\ }\href@noop {} {\bibfield  {journal} {\bibinfo  {journal} {Adv.
  Theor. Math. Phys.}\ }\textbf {\bibinfo {volume} {2}},\ \bibinfo {pages}
  {303} (\bibinfo {year} {1998})},\ \Eprint
  {http://arxiv.org/abs/q-alg/9707029} {arXiv:q-alg/9707029 [q-alg]}
  \BibitemShut {NoStop}%
\bibitem [{\citenamefont {Connes}\ and\ \citenamefont
  {Kreimer}(1998)}]{Connes:1998qv}%
  \BibitemOpen
  \bibfield  {author} {\bibinfo {author} {\bibfnamefont {A.}~\bibnamefont
  {Connes}}\ and\ \bibinfo {author} {\bibfnamefont {D.}~\bibnamefont
  {Kreimer}},\ }\href {\doibase 10.1007/s002200050499} {\bibfield  {journal}
  {\bibinfo  {journal} {Commun. Math. Phys.}\ }\textbf {\bibinfo {volume}
  {199}},\ \bibinfo {pages} {203} (\bibinfo {year} {1998})},\ \Eprint
  {http://arxiv.org/abs/hep-th/9808042} {arXiv:hep-th/9808042 [hep-th]}
  \BibitemShut {NoStop}%
\bibitem [{\citenamefont {Connes}\ and\ \citenamefont
  {Kreimer}(2000)}]{Connes:1999yr}%
  \BibitemOpen
  \bibfield  {author} {\bibinfo {author} {\bibfnamefont {A.}~\bibnamefont
  {Connes}}\ and\ \bibinfo {author} {\bibfnamefont {D.}~\bibnamefont
  {Kreimer}},\ }\href {\doibase 10.1007/s002200050779} {\bibfield  {journal}
  {\bibinfo  {journal} {Commun. Math. Phys.}\ }\textbf {\bibinfo {volume}
  {210}},\ \bibinfo {pages} {249} (\bibinfo {year} {2000})},\ \Eprint
  {http://arxiv.org/abs/hep-th/9912092} {arXiv:hep-th/9912092 [hep-th]}
  \BibitemShut {NoStop}%
\bibitem [{\citenamefont {Kreimer}(2010)}]{Kreimer:2009jt}%
  \BibitemOpen
  \bibfield  {author} {\bibinfo {author} {\bibfnamefont {D.}~\bibnamefont
  {Kreimer}},\ }\href@noop {} {\bibfield  {journal} {\bibinfo  {journal} {Clay
  Math. Proc.}\ }\textbf {\bibinfo {volume} {11}},\ \bibinfo {pages} {313}
  (\bibinfo {year} {2010})},\ \Eprint {http://arxiv.org/abs/0902.1223}
  {arXiv:0902.1223 [hep-th]} \BibitemShut {NoStop}%
\bibitem [{\citenamefont {Panzer}\ and\ \citenamefont
  {Schnetz}(2016)}]{Panzer:2016snt}%
  \BibitemOpen
  \bibfield  {author} {\bibinfo {author} {\bibfnamefont {E.}~\bibnamefont
  {Panzer}}\ and\ \bibinfo {author} {\bibfnamefont {O.}~\bibnamefont
  {Schnetz}},\ }\href@noop {} {\  (\bibinfo {year} {2016})},\ \Eprint
  {http://arxiv.org/abs/1603.04289} {arXiv:1603.04289 [hep-th]} \BibitemShut
  {NoStop}%
\bibitem [{\citenamefont {Joni}\ and\ \citenamefont {Rota}(1979)}]{JoniRota}%
  \BibitemOpen
  \bibfield  {author} {\bibinfo {author} {\bibfnamefont {S.~A.}\ \bibnamefont
  {Joni}}\ and\ \bibinfo {author} {\bibfnamefont {G.~C.}\ \bibnamefont
  {Rota}},\ }\href@noop {} {\bibfield  {journal} {\bibinfo  {journal} {Studies
  in Applied Mathematics}\ }\textbf {\bibinfo {volume} {61}},\ \bibinfo {pages}
  {93} (\bibinfo {year} {1979})}\BibitemShut {NoStop}%
\end{thebibliography}%
  
\end{document}